\documentclass[preprint,showpacs,amsmath,amssymb,12pt]{revtex4}
\usepackage{graphicx}
\usepackage{psfrag}
\begin{document}
\title{Solving the Advection-Diffusion Equations in Biological
  Contexts using the Cellular Potts Model.}  
\author{Debasis Dan} \email{ddan@indiana.edu}
\affiliation{Biocomplexity Institute and Department of Physics, 727 E.
  3rd Street, Swain Hall West 159, Indiana University, Bloomington, IN
  47405-7105 USA.} 
\author{Chris Mueller}
\affiliation{Biocomplexity Institute and Department of Physics, 727 E.
  3rd Street, Swain Hall West 159, Indiana University, Bloomington, IN
  47405-7105 USA.}
\author{Kun Chen}
\affiliation{Biocomplexity Institute and Department of Physics, 727 E.
  3rd Street, Swain Hall West 159, Indiana University, Bloomington, IN
  47405-7105 USA.}
\author{James A.  Glazier } \email{glazier@indiana.edu}
\affiliation{Biocomplexity Institute and Department of Physics, 727 E.
  3rd Street, Swain Hall West 159, Indiana University, Bloomington, IN
  47405-7105 USA.} 
\begin{abstract}
  The Cellular Potts Model (\textit{CPM}) is a robust, cell-level
  methodology for simulation of biological tissues and morphogenesis.
  Both tissue physiology and morphogenesis depend on diffusion of
  chemical morphogens in the extra-cellular fluid or matrix
  (\textit{ECM}).  Standard diffusion solvers applied to the cellular
  potts model use finite difference methods on the underlying CPM
  lattice.  However, these methods produce a diffusing field tied to
  the underlying lattice, which is inaccurate in many biological
  situations in which cell or ECM movement causes advection rapid
  compared to diffusion. Finite difference schemes suffer numerical
  instabilities solving the resulting advection-diffusion equations.
  To circumvent these problems we simulate advection-diffusion within
  the framework of the CPM using off-lattice finite-difference
  methods. We define a set of generalized fluid particles which detach
  advection and diffusion from the lattice. Diffusion occurs between
  neighboring fluid particles by local averaging rules which approximate
  the Laplacian. Directed spin flips in the CPM handle the advective
  movement of the fluid particles.  A constraint on relative velocities in
  the fluid explicitly accounts for fluid viscosity.  We use the CPM
  to solve various diffusion examples including multiple instantaneous
  sources, continuous sources, moving sources and different boundary
  geometries and conditions to validate our approximation against
  analytical and established numerical solutions.  We also verify the
  CPM results for Poiseuille flow and Taylor-Aris dispersion.

\end{abstract}
\pacs{02.70.Uu, 05.10.-a, 47.11.+j,87.10.+e} \maketitle

\section{Introduction}
   
Advection-Diffusion equations (\textit{ADE}) describe a broad range of
natural phenomena. They occur in diverse field including physics
\cite{physics}, chemistry \cite{chem}, biology, geology
\cite{atmos,geo} and even in migration and epidemiology \cite{disease}.
They describe the flow (deterministic) and the spread (stochastic) of
a density (of a chemical, heat, charge) which a fluid or deformable
solid carries. The simplest ADE is :
\begin{equation}
 \frac{\partial n}{\partial t} = D\nabla^{2}n - \overrightarrow v \cdot 
 \overrightarrow \nabla n,
\end{equation}
where $n$ is the density of the transported substance, $D$ its
diffusion constant (here assumed uniform in space) and
$\overrightarrow v$ is the velocity field. The velocity field in turn
couples to the pressure field of the medium through the Navier-Stokes
equations.  Though we can solve the problem analytically in steady
state with simple boundary conditions, most physically relevant ADEs
appear within sets of nonlinear coupled equations or with nontrivial
boundary conditions where analytical solutions are not possible
\cite{faber}. Hence a vast literature exists on how to solve ADEs.
Most solvers use either finite difference (\textit{FD}) or finite
element (\textit{FE}) \cite{fem1} schemes.  Besides these
deterministic approaches, several schemes use Lattice-Boltzmann
(\textit{LB}) methods \cite{lb} like Flekkoy's method \cite{flek},
Dawson's method \cite{daw}, the moment-propagation method
\cite{moment}.  ADEs in general are difficult to solve in the absence
of separation of diffusion and advection time scales or in the
presence of moving boundaries. Most lattice-based methods locally
refine the grid during
solution to avoid instabilities.
Moreover, explicit LB methods require time averaging of the torque to
avoid instabilities \cite{ladd}.  Hence the computational cost of both
LB and deterministic methods shoots up.  Non-staggered FD grids may
show grid-decoupling instabilities \cite{ladd}. Also, all explicit
methods require consideration of the general stability constraints
from linear analysis, most notably the Neumann diffusive criterion
linking the time step and the square of the grid size.  In this
article we try to address the problems associated with incorporating
advection-diffusion in biologically-motivated, multiscale simulations,
specifically those which use the Cellular Potts Model (\textit{CPM})
to model cell behaviors.

Diffusion of morphogens and flow of extra-cellular matrix
(\textit{ECM}) are crucial to many biological phenomena, including
wound healing, morphogenesis, \textit{e.g.}, during mesenchymal
condensation or gastrulation \cite{wei} and the immune response where
cells emerge from the microvasculature and migrate towards sites of
inflammation to kill bacteria, other pathogens and cancer. The generic
mechanisms common to all these processes are changes in cell
velocities (\textit{chemotaxis}) or/and differentiation in response to
the temporal and spatial variations of chemical morphogens.  Other
classic examples of diffusion-driven morphogenesis involve Turing
instabilities. Turing instabilities arise due to different diffusion
rates of two or more reacting chemicals resulting in competition
between activation by a slow-diffusing chemical (\textit{activator})
and inhibition by a faster chemical (\textit{inhibitor}). Pattern
formation during morphogenesis due to Turing instabilities is a
subject by itself \cite{turing}.

Besides chemotaxis, the formation and rate of extension of pseudopods
which crucially depends convective mass transport
\cite{psedupod,adapt} also influences cell motility.  Hydrodynamic
shear can also increase cell-cell
adhesion efficiency by increasing the number of binding receptors. 
Shear has a profound effect on neutrophil-platelet adhesion and
neutrophil aggregation, key events in acute coronary syndromes like
arterial thrombosis (\cite{thrombo}).
Extensive work has shown that both fluid shear amplitude and shear
exposure time modulate the interactions between polymorphonuclear
leukocytes and colon carcinoma cells \cite{pmn}. Gene expression and
protein synthesis in endothelial cells also change upon application of
arterial shear stresses \cite{genes}. In a prominent example, fluid
shear allows optimal L-selectin-mediated leukocyte rolling only above
a minimum threshold shear rate \cite{roll}.  Hence multicellular
modeling tools have to properly account for the advection-diffusion:
diffusion influencing the spatial and temporal distribution of
chemical morphogens, advection controlling the rates of cell
collisions, deformation, receptor-ligand bond formation
\cite{examples}, adhesion and enhanced mixing of chemical morphogens.


Simulations of the development of multicellular organisms take diverse
mathematical approaches: continuum FE-based models \cite{fem} of
reaction-diffusion which consider cell density as a continuous
variable \cite{reaction-diffusion}, hybrid models like E-cell
\cite{ecell} and cellular automaton approaches \cite{alber}.  Glazier
and Graner developed the cell-level CPM, an extension of the
energy-based large-$Q$ Potts model, for organogenesis simulations
\cite{first}. The basic CPM explains how surface binding energies
drive cell movement and models cell sorting from an initial random
distribution into different patterns depending on the cell adhesion
coefficients at homotypic, heterotypic and cell-medium boundaries
\cite{second}. It also provides a platform on which to build
simulations of a wide range of biological experiments by including
additional mechanisms like directed active movement due to external
fields, \textit{e.g.}  chemotaxis to a
chemical field gradient, gravity or cell polarity. 
CPM applications include modeling mesenchymal condensation
\cite{organo,stripe,morpho}, the complete life-cycle of {\it
  Dictyostelium discoideum} \cite{stan}, tumor growth \cite{tumor},
vascular development \cite{roeland}, immune response and limb growth
\cite{limb}.  Unlike the simple Turing mechanism where cells have no
feedback on the chemical field, most CPM implementations include this
feedback which can give rise to completely different patterning from
the Turing mechanism. In the CPM, patterns can arise under the
influence of a single chemical field \cite{wei} due to cell movement,
biased by gradients in cell-cell adhesion and cell-ECM binding, which
is impossible in Turing mechanism. This unique mechanism also differs
from chemotaxis, which requires long-range cell movement
\cite{stripe}.


All existing CPM implementations suffer from four main limitations:
(1) They do not include viscous dissipation explicitly. Instead
dissipation arises from the Metropolis-Boltzmann energy-minimization
dynamics. This implicit dissipation makes viscosity hard to calculate
or control.  (2) They do not explicitly describe force transduction
through cells, which arises through the volume constraint and surface
constraint (if used) of the Hamiltonian.  (3) Aggregates of cells
modeled with an ordinary CPM Hamiltonian are highly overdamped.
Modeling the ECM or fluid as an array of generalized cells using the
normal CPM produces a flow resembling the overdamped flow of
biological fluid but this fluid slips at solid surfaces and exerts no
shear force unlike a normal fluid. An alternative approach which
describes the fluid as a single, large, unconstrained generalized cell
produces nonlocal movement.  Moreover, it cannot advect chemicals or
transmit shear forces.  (4) The CPM has no intrinsic concept of
rigid-body motion. We will describe an algorithmic solution to this
last problem, which makes the CPM look more like a FE simulation, in a
future paper.

All of these problems result from the CPM spins being tied to the
underlying lattice, rather than to objects they describe.  The
solution in each case is to adopt an FE approach suited to the CPM,
which takes the behavior off-lattice. Applying standard off-lattice
methods in the CPM has inherent problems. Since in the CPM cell
movement occurs by boundary fluctuations, connecting normal FE
fluid-solvers to the CPM at surfaces requires local grid refinement
during cell movement to avoid numerical
instabilities. 
Hence its computational cost is high.

To introduce advection-diffusion in the CPM correctly and efficiently,
we propose an off-lattice scheme consistent and in harmony with the
CPM algorithm. We subdivide the ECM into small fluid particles having the
normal properties of generalized cells like differential adhesivity, a
surface constraint and a volume constraint. The fluid particles carry
chemical morphogens and we assume the concentrations are uniform
inside them.  Diffusion occurs between neighboring fluid particles by
local averaging rules which approximate the Laplacian. Spin flips in
the CPM occur only at the boundaries of the cells or particles. A force on the
fluid in any direction due to pressure gradients or external forces
biases the probability of spin flips and generates directed motion
\cite{yi_thesis}.  
We introduce viscosity into the CPM by explicitly in the Hamiltonian
including a relative-velocity constraint between the neighboring fluid
particles.  This scheme allows us to solve the ADEs and generates the
creeping flow of a highly viscous fluid.

   
In most biologically relevant regimes (\textit{e.g.}, \textit{E. Coli}
in water), we encounter low Reynolds number (\textit{Re}) flow $\sim
10^{-5}$, with the typical diffusion coefficient of chemical
morphogens being $\sim 10^{-4} \mu m^{2}/s$.  Thus the Peclet number
(defined as $\frac{R v}{D}$, where $R, v, D$ are typical size of the
system, the typical fluid velocity and diffusion coefficient
respectively) is as low as $\sim 10^{-2}$ \cite{life}.  We explicitly
exclude blood flow in the circulatory system , where Re numbers (hence
inertia) can be quite large and where specialized solution techniques
already exist \cite{blood}. Since most CPM simulations do not demand
high precision, sophisticated methods like mimetic finite difference
\cite{mimetic} become a computational bottle-neck without many
advantages.  On the other hand, our ADE scheme is very stable, with
spatial resolution equal to the mean diameter of the fluid particles,
which are much smaller than the simulated biological cells.  Our
scheme seamlessly integrates into the main Monte-Carlo loop of the CPM
simulation and boundary conditions like absorbing boundaries or
no-flux boundaries are simple to implement.  More-over our off-lattice
scheme does not require re-meshing.  We discuss these issues in detail
below.

This paper mainly focuses on the validity and utility of the CPM ADE
solver.  Section $2$ briefly describes the CPM along with the
diffusion and advection scheme. Section 3 discusses various cases,
including diffusion from two point sources, a continuous source and a
moving source with either reflecting or absorbing boundary conditions
and the flow profile in Poisueille's flow.  Section 4, outlines future
directions in developing the ADE scheme.  We will address in a future
paper certain additional conditions, flow with inertia, effects of low
or high Peclet numbers, constitutive properties of the fluid phase,
\textit{etc}.  As we have stated before, ours method provides
flexibility and efficiency in the biologically relevant regime of low
Peclet and Reynolds numbers and where high numerical accuracy is not
crucial.

\section{The Model}

The Potts model is an energy-based, lattice Cellular-Automaton
(\textit{CA}) model equivalent to an Ising model with more than two
degenerate spin values.  We typically use a cubic lattice with
periodic or fixed boundary condition in each direction. We use $3$rd
or $4$th nearest neighbor interactions to reduce lattice-anisotropy
induced alignment and pinning. Each lattice site in the Potts model
has a spin value. The energy, or \textit{Hamiltonian} sums the
interaction energies of these spins, according to predefined rules. In
single-spin dynamics (like Metropolis dynamics) the spin lattice
evolves towards its equilibrium state by minimizing the interaction
energy through spin flips.  Multi-spin dynamics like Kawasaki dynamics
are also possible. Though the original Potts model studies focuses on
equilibrium properties, it can also model quasi-equilibrium dynamical
properties \cite{barkema}. Using deterministic schemes for spin flips,
patterns often stick in local minima. Finite-temperature Monte-Carlo
schemes circumvent this problem. These schemes accept a spin flip with
temperature dependent Boltzmann or modified Boltzmann probability if
the configuration encounters a potential barrier (a greater energy
after the spin flip than before) \cite{barkema}. 

We pick a target lattice site at random and one of its alien neighbor,
also selected at random and attempt to flip the target spin to the
value of the selected neighbor.  In the modified Metropolis algorithm
we employ, if the spin flip would produce a change in energy $\Delta
H$, we accept the change with probability $P$ given by
\begin{equation}
  \label{eq:prob}
  P(\Delta H) = \left\{ \begin{array}{ll}
                         \exp(-\Delta H/T)  & \mbox{ if } \Delta H > 0, \\
                         1                                & \mbox{ otherwise, }
                         \end{array}
                         \right.
\end{equation}
where $T$ is the fluctuation temperature. $T$ controls the rate of
acceptance of the proposed move. For very large $T$, all the moves are
accepted and the dynamics is a random walk in the absence of barriers,
\textit{i.e.} interaction energies included in the Hamiltonian are
effectively zero producing a disordered phase.  For very small values
of $T$ the dynamics is deterministic and can trap in local minima. We
choose $T$ as the median value of the distribution of $\Delta H$,
which is below the order-disorder phase transition temperature. All of
our results are very robust with respect to variation of $T$. S. Wong 
has recently shown that optimizing the dynamics of the modified 
Metropolis algorithm requires changes to the acceptance probability 
in eqn.~\ref{eq:prob} \cite{skai}. However, we do not implement these
changes here. Our unit
of time is Monte-Carlo sweep (\textit{MCS}), where $1 MCS = L^{3}$ spin 
flip attempts, $L$ being the system size.

\begin{figure}[h]
  \includegraphics[scale=0.5]{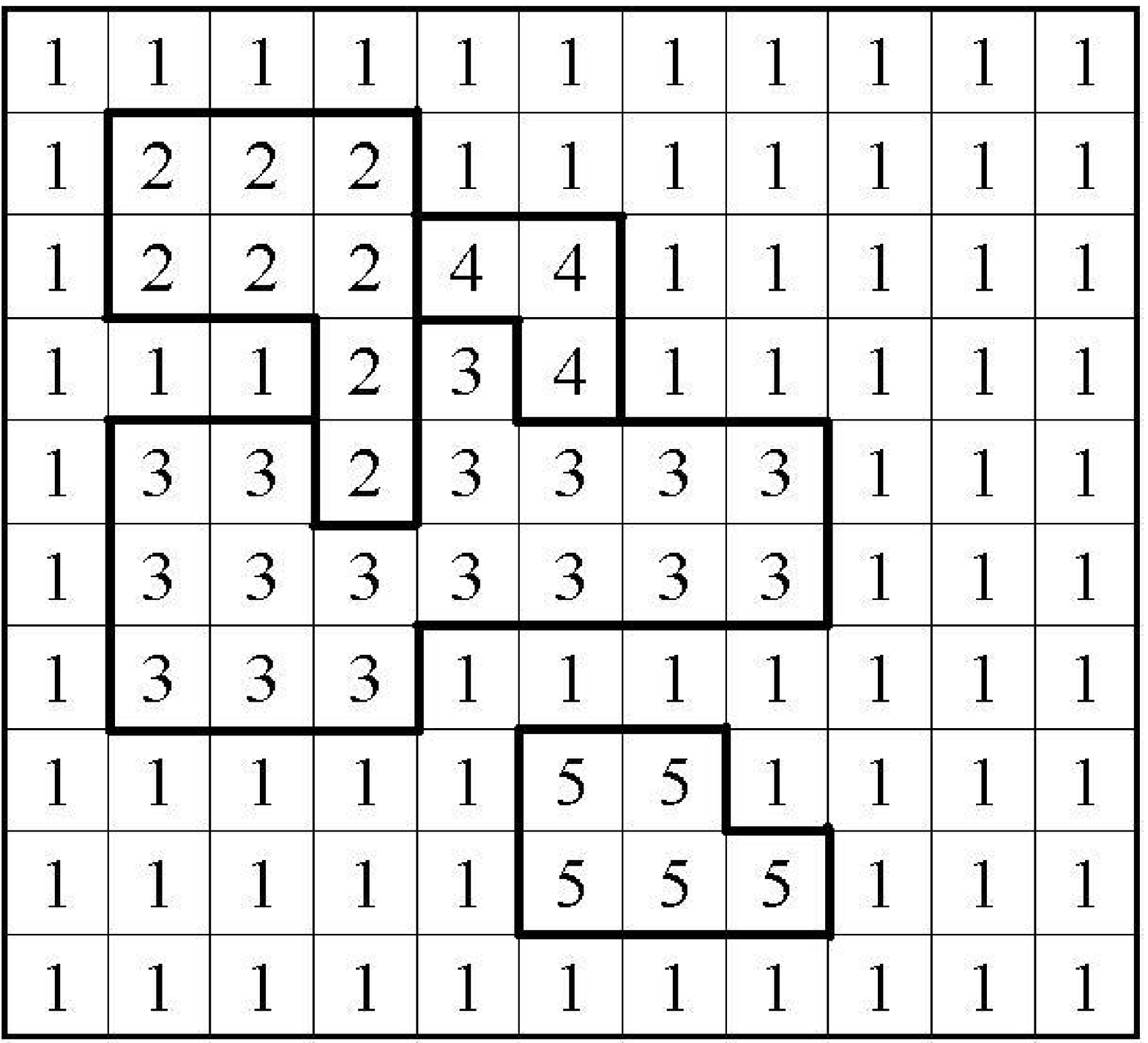}
  \caption{A typical cell configuration in CPM. The bold lines denote 
    cell boundary. }
  \label{lattice}
\end{figure}

The CPM adapts the Potts model to the context of biology. A CPM cell
is a collection of lattice sites with same spin value (or index)
$\sigma_{i}$. Each cell has a unique spin $\sigma$ (see
fig.~\ref{lattice}). Cells may also have additional
characteristics,\textit{e.g.}, a type $\tau$.  Links between different
sites with spins define cell boundaries.  So cells have both volume
and a surface area. The volume, area, radius relation is highly
non-Euclidean for small cells.

The CPM Hamiltonian contains a variable number of terms.  The
interaction between pairs of biological cells involves an adhesive or
repulsive interfacial energy.
This interfacial energy is precisely the Potts energy, the sum of the
interactions of neighboring unlike spins, across a link.  Each
mismatched link contributes a cell-type dependent binding energy per
unit area $J(\tau,\tau^{'})$, where $\tau \mbox{ and } \tau^{'}$ are
the type of cells on either side of the link.  In the CPM the
effective cell-cell interaction energy is:
\begin{equation}
E_{adhesion} = \sum_{(i,j,k),(l,m,n) neighbors}J(\tau_{\sigma(i,j,k)},\tau_{\sigma(l,m,n)})
              (1-\delta_{(\sigma(i,j,k),\sigma(l,m,n))}),
\end{equation}
where $\delta_{(\sigma^{'},\sigma)} = 1 \mbox{ if } \sigma^{'}=\sigma,
\mbox{ otherwise } \delta_{(\sigma^{'},\sigma)} = 0.$

At any time $t$, a cell, of type, $\tau$, has a volume $v(\sigma,t)$
and surface area $s(\sigma,t)$.  The volume is simple to define,
$v(\sigma_{0}) = \sum_{i,j,k} \delta_{(\sigma_{0},\sigma(i,j,k))}$,
whereas surface area is more complex, since it depends on the
interaction range of the lattice, $s(\sigma_{0}) =
\sum_{i,j,k}\delta_{(\sigma_{0},\sigma_{i,j,k})}\sum_{i^{'},j^{'},k^{'}}
(1 -\delta_{(\sigma(i,j,k),\sigma(i^{'},j^{'},k^{'}))})$, where
$(i,j,k) \mbox{ and } (i^{'},j^{'},k^{'})$ are neighboring lattice
sites.  Each cell has an effective volume elasticity, $\lambda_{v}$
and target volume $v_{target}(\sigma,t)$. Larger values of
$\lambda_{v}$, produce less compressible cells. We typically choose
$v_{target}^{2}\lambda_{v} >$ other constraint energies. This
compressibility makes little difference in low Reynolds number flow
but makes pattern evolution less stiff.
We also define a membrane elasticity, $\lambda_{s}$, and a target
surface area $s_{target}(\sigma,t)$ to maintain the generalized shape
of the cells.  The energy contributions due to surface and volume
fluctuations are:
\begin{equation}
E_{surface} = \lambda_{s}(s(\sigma,t)-s_{target}(\sigma,t))^{2}, 
\end{equation}
\begin{equation}
E_{volume} = \lambda_{v}(v(\sigma,t)-v_{target}(\sigma,t))^{2}.
\label{volume_term}
\end{equation}

We can extend the Hamiltonian to include a uniform external force
$\vec{F}$, acting on all cells by including the term:
\begin{equation}
E_{force} = -\sum_{(ijk),(lmn) neighbors} \vec{\bf{F}} \cdot 
\vec{\bf{r}}_{i,j,k}(1-\delta_{\sigma(i,j,k),
\sigma(l,m,n)}),
\end{equation}
where $\vec{\textbf{r}}_{ijk}$ is the position vector at the lattice site
($i,j,k$).

Previous CPM applications often treated fluid or ECM as a single large
cell with no constraints. Here we take coarse-grained approach to
describe ECM. We assume the ECM consists of hypothetical fluid cells
(which we call particles to avoid confusion with the modeling of
biological cells) having all the characteristic interactions and
constraints of regular CPM cells.  The volume constraint and the
surface tension determine the elastic nature of the fluid.  The fluid
particles can move with respect to one another like regular CPM cell
via spin flips. Thus, local pressure developed due to movement or
enlargement of actual biological cells will translate into motion of
the surrounding ECM.  This fluid motion causes advection and mixing
along with molecular diffusion of chemical morphogens. We restrict
consideration to the highly overdamped viscous world that most
biological cells experience, so our fluid particles lack inertia.  We
also restrict to situations where the velocity of movement is much
less than the velocity of
sound, which is one lattice unit per MCS.  
 
We now introduce a relative velocity constraint between the
cells/particles which faithfully captures the effects of shear due to
the viscosity of the medium. In the CPM, velocity is a cell property
defined as the displacement of the center of mass of the cell per MCS.
Since the velocity gradient terms in the direction of the flow
(\textit{e.g.}, $\frac{\partial u_{i}}{\partial x_{i}}$) are the rate
of change of volume \cite{landau} which eqn. \ref{volume_term} already
includes, we need to keep only the contributions of cross terms of the
form $ (\frac{\partial u_{i}}{\partial x_{j}} )^{2}$. In an
incompressible fluid ($\nabla \cdot \mathbf{u} = 0$) the cross terms
are the
dissipation energy per unit volume \cite{faber}. 
In the CPM we model this term as :
\begin{align}
E_{viscosity} & = \lambda_{viscosity}\sum_{i}\sum_{j } 
S_{ij}\frac{(V_{i_{x}} 
- V_{j_{x}})^{2}}{d_{ij}^{2}} \sqrt{\frac{(y_{i}-y_{j})^{2}+(z_{i}
-z_{j})^{2}}{d_{ij}^{2}}} \nonumber \\ 
& + \mbox{ cyclic permutation of } (x,y,z) ,
\end{align}
where the $j$'s are the indices of cells neighboring the $i$th cell
and $d_{ij} = \sqrt{(x_{i}-x_{j})^{2} +
  (y_{i}-y_{j})^{2}+(z_{i}-z_{j})^{2}}$ is the distance between the
centers of cell $i$ and cell $j$. $V_{i_{x}}$ is the $\widehat
x$-component of the velocity of the $i$th cell.  Since the cells are
of irregular shape, we further weight the energy penalty by the cells
common contact area $S_{ij}$. We ensure that the cells are simply
connected by using local connectivity checks during spin flip
attempts. $\lambda_{viscosity}$ corresponds to the viscosity
coefficient $\eta$ in the Navier-Stokes equations. $\eta$ has
dependence on other system parameters like $J, \lambda_{volume} \mbox{
  and } \lambda_{surface}$.
 

The net Hamiltonian including fluids is then :
\begin{equation}
H = E_{adhesion} + E_{surface} + E_{volume} + E_{viscosity}.
\end{equation}

\subsection*{Diffusion Scheme}

Since the motion of the fluid particles takes care of advection, we need
only to solve the diffusion on the current fluid particles configuration;
$\frac{\partial C(\vec{x},t)}{\partial t} = D \nabla^{2}
C(\vec{x},t)$, where we have assumed that the diffusion constant $D$
is constant and isotropic. Including an anisotropic or spatially
varying $D$
is a trivial extension of our method. 
We assume that the fluid particles carry chemical morphogens, whose
distribution is uniform over a given fluid particle. Equivalently we
can associate the chemical concentration with the center of mass of
the particles and think of the diffusion as taking place between them (a
FE view).  Because the shape and volume of the fluid particles are
irregular, their centers of mass do not correspond to lattice points.
We then need a numerical scheme for diffusion among fluid particles.
Few existing algorithms solve diffusion on a random or irregular
lattice.  The more sophisticated and accurate ones are computationally
expensive \cite{mimetic}.  We use a naive iterated Euler method, which
is fast and stable and reproduces different biological experiments
with good qualitative and fair quantitative accuracy.  We could of
course, use a more elaborate scheme, if necessary. We locally average
the concentration among the particles nearest neighbors. The particle
neighbors change as the fluid flows.  We approximate the Laplacian
$\nabla^{2}C(\vec {\textbf{r}},t)$ (where $C(\vec {\textbf{r}},t)$ is
the chemical concentration and $\vec{\textbf{r}}$ is the center of
mass coordinate of a fluid particle) by \cite{murray}:
\begin{equation}
  \label{eq:laplace}
  D\nabla^{2}C_{j}(t) \sim D\sum_{i \mbox{ next to }j}{\frac{ (C_{i}(\vec{\textbf{r}},t) - C_{j}(\vec{\textbf{r}},t))}{R^{2}}},
\end{equation}
where $R$ is the average radius of the fluid particles assuming them
to be spherical ($R = \frac{\sum_{i=1}^{N} V^{1/3}_{i}}{N}$) and
$C_{i}(\vec{\textbf{r}},t)$ is the concentration in nearest neighbor
fluid particles.  We use $R^{2}$ instead of $|\vec{\textbf{r}_{i}} -
\vec{\textbf{r}_{j}}|^{2}$ to avoid Neumann instability.  We can
update the concentration once or multiple times per MCS depending on
the diffusion coefficient of the chemical morphogen. $D$ along with
number of concentration updates per MCS controls the diffusion
constant. For larger diffusion coefficients we update the diffusion
step multiple times per MCS.  For typical chemical morphogens like
cAMP the diffusion constants are $\sim 10^{-4} \mu m^{2}/s $ ,
corresponding to a spread of $3\sqrt{2.\mbox{x}10^{-4}} \mu m$ $(0.04
\mu m)$ per MCS. The typical lattice spacing in our CPM corresponds to
$0.1-1.0 \mu m$, hence a fluid particle has a width of $\sim 0.3-3.0
\mu m$. Therefore a very small amount chemical needs to diffuse among
adjacent fluid particle per MCS.  In fact, for most practical purposes
we need at most one diffusion step per five to ten MCS.  Both
reflecting and constant concentration (zero concentration is
absorbing) boundaries are simple to implement.  For reflecting
boundary condition we exclude from the exchange of concentration
(eqn.~\ref{eq:laplace}) any particle whose boundary is reflecting. For
absorbing boundary condition (constant concentration), we define the
particles whose boundaries are absorbing to always have zero
concentration. The algorithm handles moving boundaries automatically
and conserves chemical concentration, unlike many other methods.

\section{Results and Discussion}
\subsection{Poiseuille Flow}
\label{sec:Pois}

\begin{figure}[h]
  \includegraphics[scale=0.75]{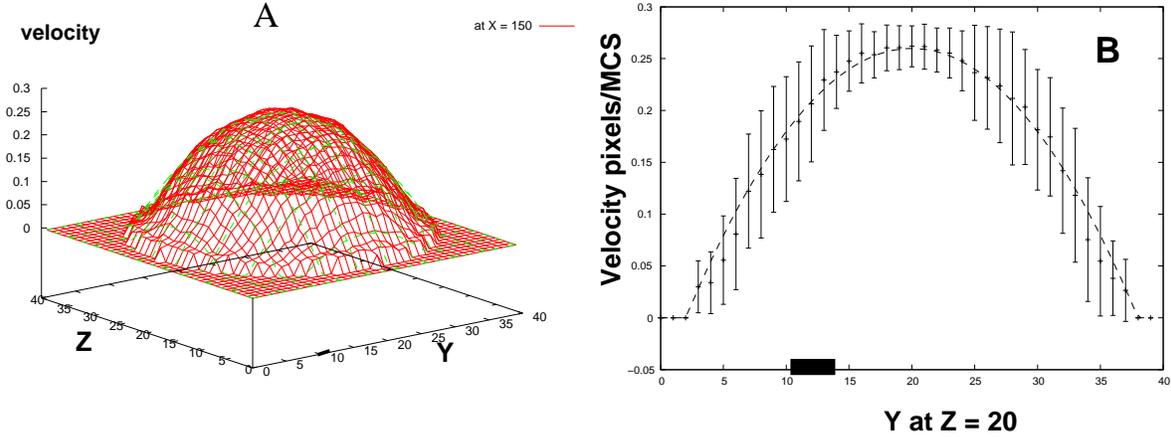}
  \caption{(A) Velocity profile in a cylindrical flow under gravity and 
    its fit to the analytical solution. (B) Velocity profile across a
    diameter of the tube with error bars, the bold segment along the
    $\widehat x$ direction denotes the width of a fluid particle.}
  \label{profile}
\end{figure}

We first discuss simulating Poiseuille flow of a viscous fluid in a
cylindrical tube with rigid walls under a uniform force field (body
force) using the CPM. We consider a cylindrical tube with a circular
cross-section of radius $18$ lattice units and length $200$ lattice
units with periodic boundary condition along the length. Each fluid
particle has an average volume of $64$ lattice points. Our CPM AD
scheme works for small forces when the fluid particles remain simply
connected. Fig.~\ref{profile} plots the ensemble averaged (50
different initial configurations) cross-sectional profile of the
viscous flow at $x = 150$ and $t=2000 \mbox{ MCS }$ with a small force
$\vec{F} = 0.05 \widehat x$ applied on all the fluid particles. The
flow is in the $x$-direction only.
\begin{figure}[h]
  \includegraphics[scale=0.5]{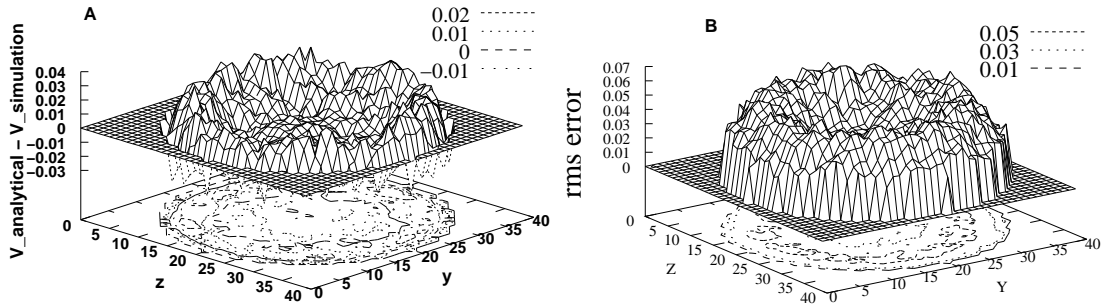}
  \caption{(A) The absolute error between the analytical solution $V_{analytical}$ and 
    the CPM simulation $V_{simulation}$ at $x$ = $150$.  (B) The rms
    error of over 50 ensembles.}
  \label{error1}
\end{figure}
The profile is parabolic as expected. We also show its fit to the
analytical solution $V(r) =
\frac{F}{4\eta}R^{2}(1-(\frac{r}{R})^{2})$, where $\eta$ is viscosity,
is a fitting parameter, $R = 18$ the radius of the cylinder and $r$ is
the distance from the axis of the cylinder. What is surprising is the
excellent agreement (an error of $<5\%$ with the analytical result
despite the coarseness of the simulation (only $9$ particles across the
diameter). All lattice points inside a fluid particle have the same
velocity, so regions where the shapes of the fluid particles are
regular for most ensembles will produce plateaux in the velocity
profile.  Since in the CPM, the velocity of fluid particles is the center
of mass velocity, a lattice point touching a boundary wall will have a
small nonzero velocity as the center of mass of the corresponding particle
lies in the interior (slip boundary).  Since the energy contribution
of the constraints near a boundary wall dominates the external applied
force and the Monte-Carlo temperature, boundary particles are more regular
in shape than interior particles. This regularity holds in all ensembles,
hence the rms error in the velocity near a boundary wall is small as
shown in fig.~\ref{error1}B and the velocity just near the boundary in
fig.~\ref{profile} has a small plateau. Reducing the size of the fluid
particless compared to the typical length scales of the flow reduces these
anomalies.
Increasing $\lambda_{viscosity}$ increases the viscosity coefficient
$\eta$.  A future paper will study the relation between $\eta$ and
$\lambda$, Monte-Carlo temperature, fluid particle size, \textit{etc}.
We shall also show additional biologically relevant hydrodynamic
flows.

We next verify our diffusion scheme under various biologically
relevant conditions.  CPM models using the modified finite temperature
Metropolis algorithm have diffusion due to the movement of the CPM
cells themselves which adds to the diffusion of chemical morphogens.
However for most CPM simulations, \textit{e.g.} a temperature $0.1$
and other parameter values used through out this paper, the diffusion
coefficient of the CPM cells is $\sim 10^{-4} pixel^{2}/MCS$, which is
much smaller than the typical diffusion coefficient of $\sim 0.1$,
that we treat in this paper. Hence for pure diffusion in a static
medium the fluid is effectively fixed. Besides the concentration
profile of the diffusing chemical in a medium, diffusion in the
presence of boundaries and moving source is crucially important in
biology.  We show that the results of our simple CPM diffusion from
CPM agree very well with corresponding analytical calculations or
finite element results. We also briefly discuss the validity of our
method for diffusion in Poiseuille flow.

The four cases we discuss below employ fluid particles with a target
volume $v_{target}=27$, unless we mention otherwise. The lattice has
$100\mbox{x}100\mbox{x}100$ sites.  We equilibrate and quench to
remove any disconnected cells which the finite-temperature
equilibration produces \cite{second}.  In our model a chemical source
at a given lattice point gives all lattice points which belong to the
fluid particle containing that chosen point the same concentration. We
apply one diffusion step per MCS. The first three cases are for static
fluids.

\subsection{Two Sources with Reflecting and Absorbing Boundaries}
\begin{figure}[h]
  \includegraphics[scale=0.5]{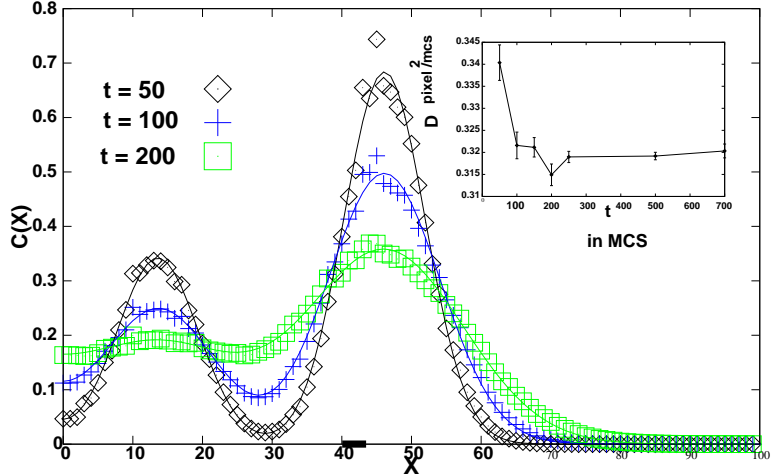}
  \caption{Symbols denote chemical profile projected on the $x$-axis 
    (summing in the Y and Z directions) at $t=50$ MCS, 100 MCS, 200
    MCS for reflecting barriers at $x=0 \mbox{ and } 100$ denoted by
    symbols. The solid line is a fit to the exact solution using the
    fitting parameter $D$. The inset shows variations in $D$ with
    time, due to coarse graining and the initial sharp distribution.}
  \label{diffu1}
\end{figure}

\begin{figure}[h]
  \includegraphics[scale=0.5]{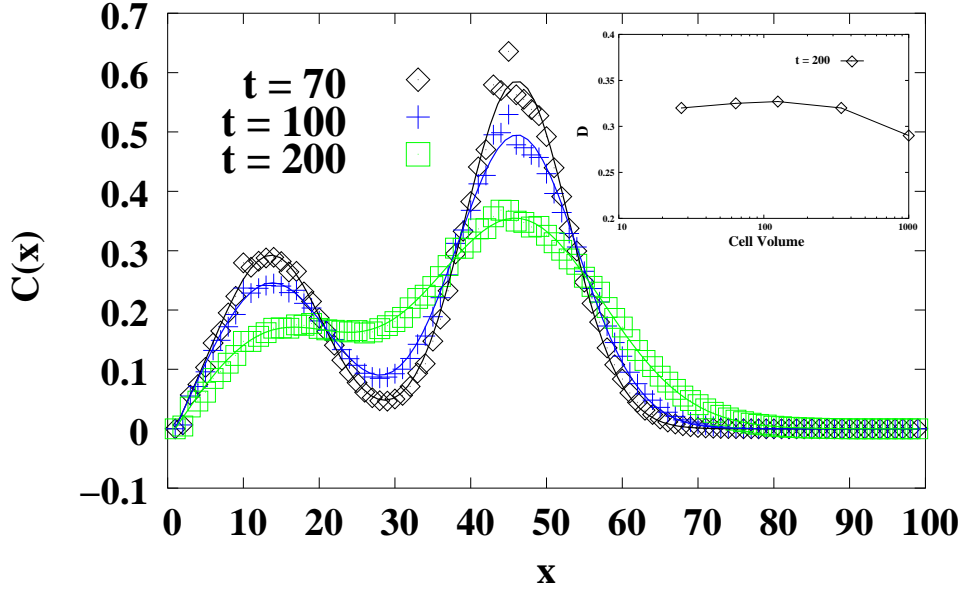}
  \caption{Chemical profile projected on the  $x$-axis from diffusion of two instantaneous 
    point sources for an absorbing barrier at $x=0$ and $x=100$ for $t
    = 70 \mbox{ MCS }, 100 \mbox{ MCS }, 200 \mbox{ MCS }$.  The inset shows the variation of $D$
    with fluid particle volume on a log scale.}
  \label{absorb}
\end{figure}

We consider two point sources at $15,50,50$ and $50,50,50$ with
initial concentrations (at $t=0$) of $5$ and $10$ respectively. The
chemicals diffuse from these instantaneous sources.  The bounding
planes of the cube are reflecting. Figure \ref{diffu1} plots the
corresponding one-dimensional diffusion profile projection (with no
ensemble average) after elapsed times $t=50$ MCS, $100$ MCS, $200$
MCS and fitted to the exact solution $C(x,t) = \frac{10.0}{\sqrt(4\pi
  D t)}(\exp^{\frac{-(x-x1)^{2}}{4Dt}} + \exp^{\frac{-(2
    L1-x-x1)^{2}}{4Dt}}+ \exp^{\frac{-(2 L2-x-x1)^{2}}{4Dt}}) +
\frac{5.0}{\sqrt(4\pi D t)}(\exp^{\frac{-(x-x2)^{2}}{4Dt}} +
\exp^{\frac{-(2 L1-x-x2)^{2}}{4Dt}}+ \exp^{\frac{-(2
    L2-x-x2)^{2}}{4Dt}}) $, $D$ being a fit parameter.  Here $L1
\mbox{ and } L2$ are the coordinates of the two reflecting boundaries
and $x1 \mbox{ and } x2 $ are the coordinates of the instantaneous
sources.  The diffusion profile matches matches very well at all
times, even near the boundaries.  The maximum relative error (compared
to the exact solution) at $t = 50 MCS$ is $5 \%$, again, surprisingly
good for such a coarse simulation. We discuss errors in detail for
absorbing boundary conditions in the next paragraph.  The inset shows
the variation of diffusion coefficient calculated from fitting to the
exact solution as function of time.  Though the diffusion constant
should remain constant with time, we observe a $5\%-6\%$ variation
from the asymptotic value at \textit{small times} due to the sharp
initial distribution producing structures smaller than the fluid particle
scale of our coarse-grained scheme. We also studied the variation of
the diffusion coefficient $D$ as function of the target volume of the
fluid particles. $D$ is constant for variations over one order of
magnitude of the target volume of the fluid particles as shown in the
inset to fig.~\ref{absorb}.  Fig.~\ref{absorb} also shows the chemical
profile for an absorbing barrier at $x=0$ and $x=100$. The figures
show that the diffusion coefficient for both reflecting and absorbing
barriers cases is same.  Hence for a wide range of fluid particle
volumes, our CPM ADE algorithm faithfully reproduces diffusion from
two point sources.

\begin{figure}
  \includegraphics[scale=0.5]{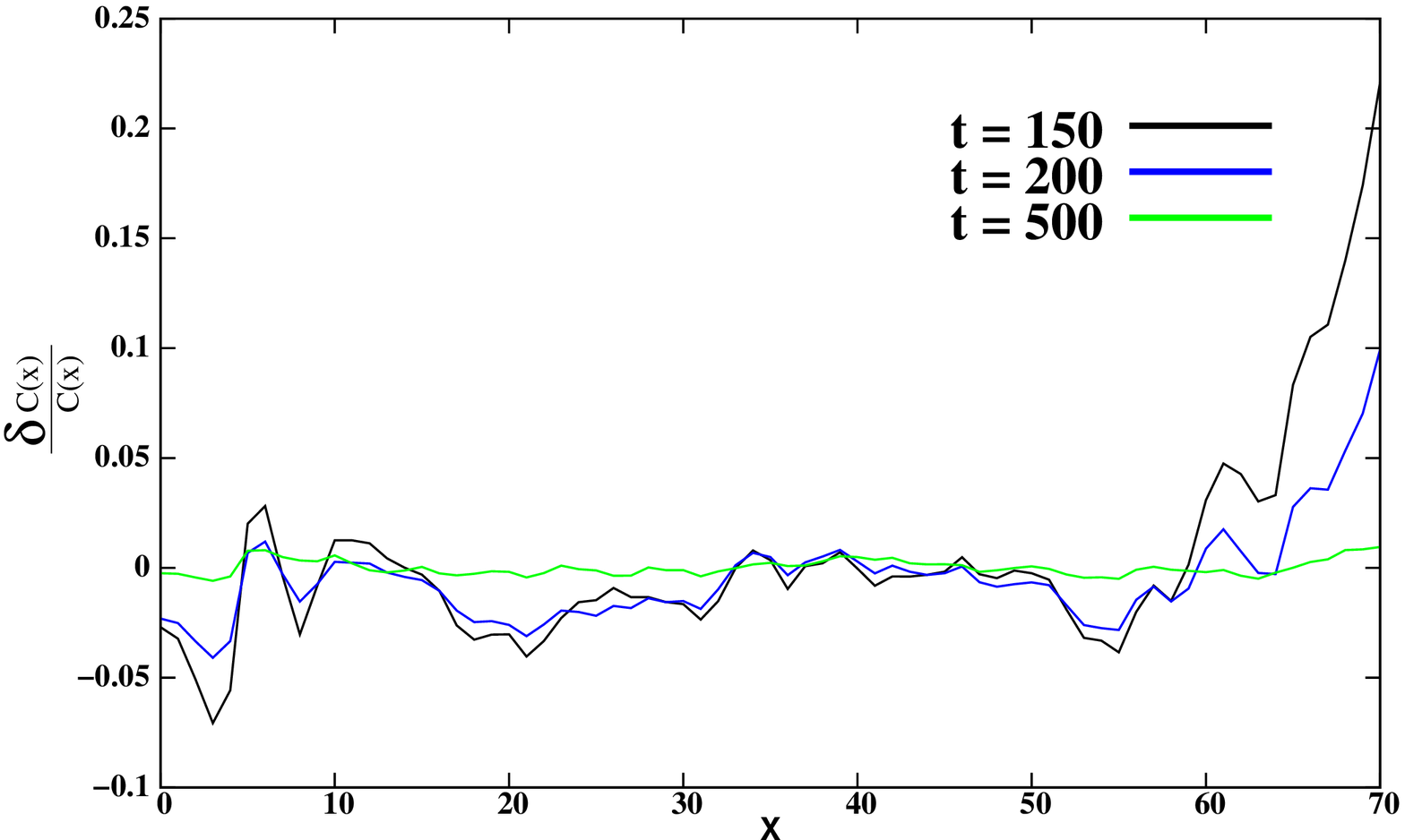}
\caption{Relative error along the $x$ lattice direction for an absorbing 
  barrier.}
  \label{error}
\end{figure}

Fig.~\ref{error} plots the relative error as a function of position at
different times. As mentioned in the last paragraph the sharp
distribution at the initial time produce errors large compared to
later time, \textit{i.e.}, if we ignore the large errors in the tail
(as the magnitude of concentration is extremely small in the tails)
the relative error at $t=150$ is $< 5\%$ whereas at $t = 500$ it is $<
0.5 \%$.  In the actual biological situation, cells secrete chemical
morphogens over their whole membrane surface and hence such singular
cases of high point concentration of chemical rarely occur.

\subsection{Two Sources with a Reflecting Obstacle inside the Medium}

Since biological cells can be impermeable to many chemical morphogens,
they can act as reflecting boundaries within the fluid medium. We
check this situation for the simple test case of a cylindrical barrier
with rectangular cross-section ($15\mbox{x}15$, pixels centered at
($50,50,50$) and the axis along the $x$-direction) within the fluid
medium.  As in the previous case, we place two point sources near the
two opposite corners of the rectangle at ($50,40,40$) and
($50,60,60$).
 \begin{figure}[h]
   \includegraphics[scale=0.5]{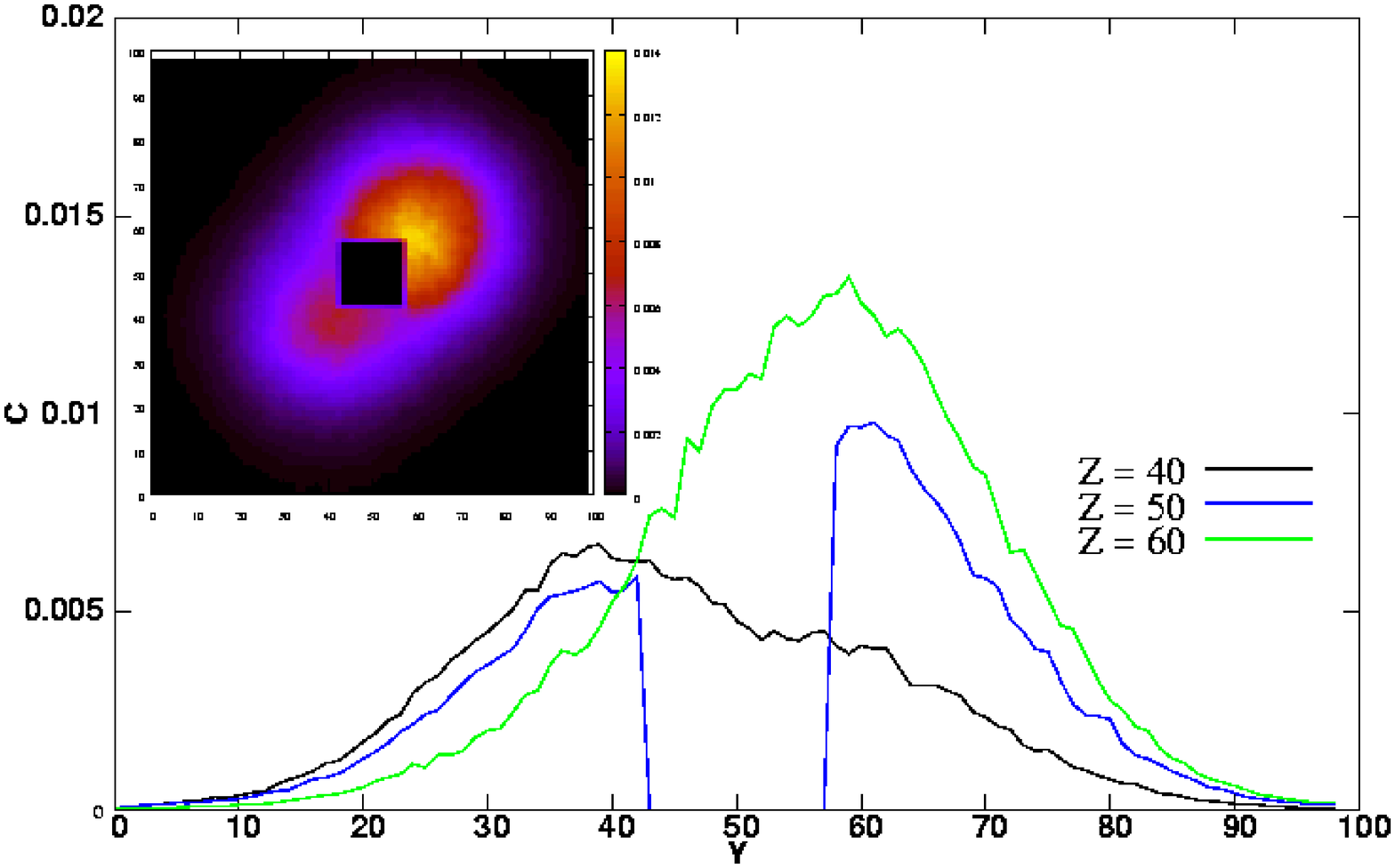}
   \includegraphics[scale=0.5]{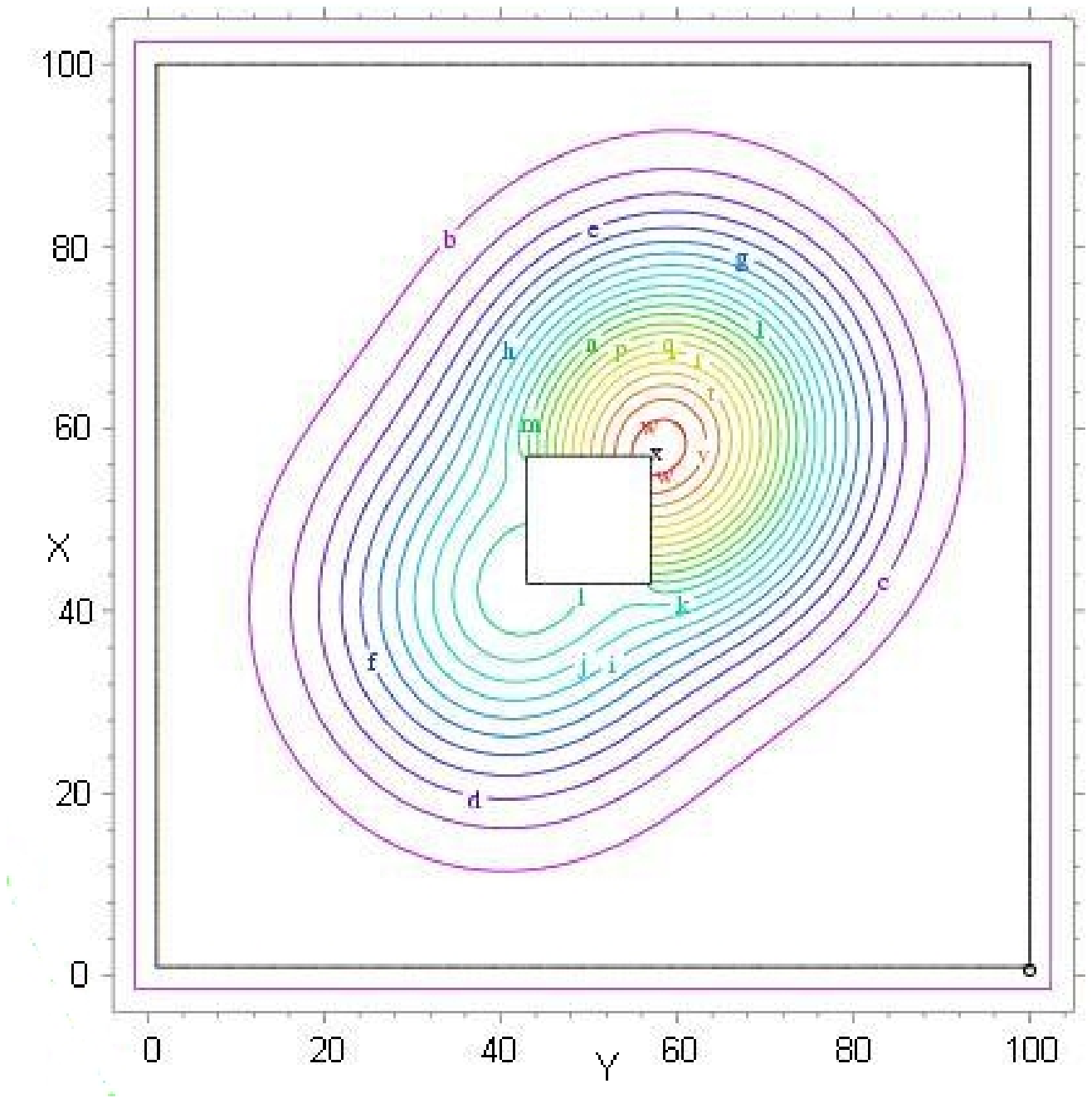}
   \caption{Chemical profile cross-section along $z=40,50 \mbox{ and }60$  in the 
     presence of a cylindrical reflecting barrier with axis along the
     $\widehat x$ direction and rectangular cross-section. The sources
     with initial concentration $5 \mbox{ and }10$ are placed at
     $50,40,40$ and $50,60,60$. The inset shows 2d projection of the
     profile.  The lower figure shows a two-dimensional projection
     obtained from finite element calculations. The concentration
     decreases as we go away from the center contours (yellow lines).}
   \label{diffu2}
 \end{figure}
 We recover the same diffusion coefficient as in the previous case.
 Fig.~\ref{diffu2} plots the one-dimensional cross-section of the
 diffusion profile at three different $z$ positions. Along $z = 40
 \mbox{ and } 60$ the barrier is absent and the diffusion is merely
 the superposition of that from the two sources. Along $z = 50$ we see
 the effect of the reflecting barrier. The inset on the right side
 shows the concentration profile obtained from a finite element
 calculation for this situation which matches very well with the CPM
 concentration profile in the left inset.

\subsection{Moving Continuous Source}
Moving cells often secrete morphogens.  Hence we correctly simulate
cells' chemotactic response to secreted chemicals only if we
faithfully reproduce diffusion from moving sources. To test our
simulation we assign an arbitrary fluid particle a constant
concentration $C_{0}$ (continuous source) and uniform velocity $v$
along the $x$ direction.  We keep the source sufficiently distant from
the boundary to avoid boundary effects. At $t=0$ the source is at
$x=35$ and moves with velocity $u=0.04$ pixel/MCS.
We fit the CPM chemical profile projection in the $x$ direction (no
ensemble average) with the 1D analytic solution:
\begin{eqnarray}
  \label{eq:moving}
  \alpha(x) &=& ((x-x_{0})^{2})/(4D), \\
  \beta &=& u^{2}/(4D),  \\
  \gamma(x) &=& \exp((x-x_{0})u/(2D)), \\
  C(x,t) &=&  \frac{C_{0}}{2}\gamma(x)(\exp(2\sqrt {\alpha(x)\beta })
  Erfc(\sqrt{\frac{\alpha(x)}{t}}+\sqrt{\beta t}) \\
         & &    + \exp(-2\sqrt{\alpha(x)\beta})Erfc(\sqrt{\frac{\alpha(x)}{t}}
            -\sqrt{\beta t})).
\end{eqnarray}

\begin{figure}[h]
  \includegraphics[scale=0.5]{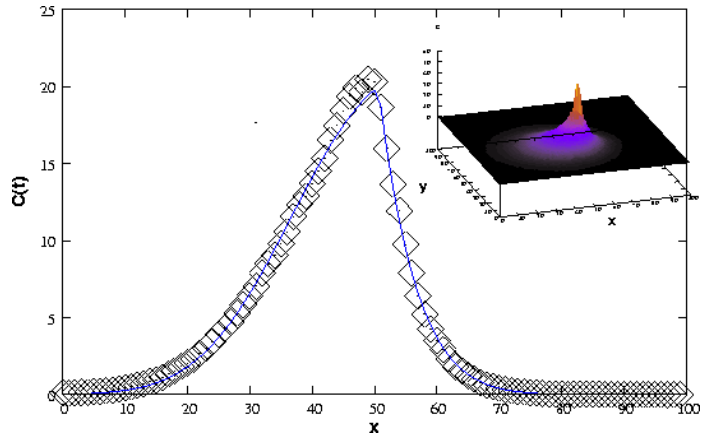}
  \caption{CPM simulation of the chemical profile from a moving source 
    (the other two coordinates integrated out) at $t=300$ MCS. The
    solid line denotes the fit to eqn.~\ref{eq:moving} .  The inset
    shows a two dimensional projection of the simulation, where red
    denotes the highest chemical concentration and black the lowest.}
  \label{diffu3}
\end{figure}
Figure~\ref{diffu3} shows that the CPM diffusion agrees very well with
the analytical solution. The diffusion coefficient obtained from the
fit is $D = 0.29$.
\subsection{Taylor-Aris Dispersion in Poiseuille flow}
\begin{figure}[h]
  \includegraphics[scale=0.5]{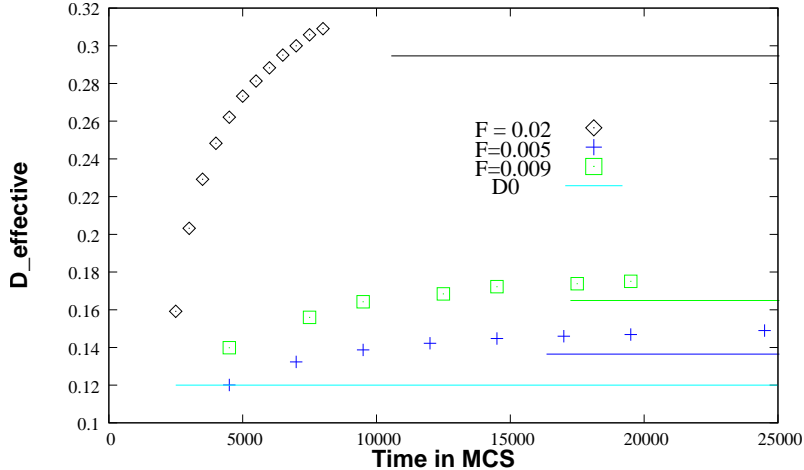}
  \caption{Effective diffusion coefficient for mean flow velocities $\overline{v} = 0.056,
    0.028 \mbox{ and } 0.016$ respectively from top to the bottom
    curve. The solid lines show the analytical results corresponding
    to above mean velocities.}
  \label{fig:dispersion}
\end{figure}

\begin{figure}[h]
  \includegraphics[scale=0.75]{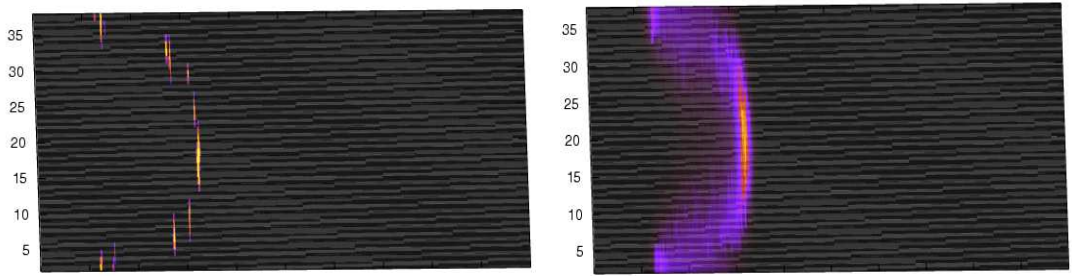}
  \caption{Cross-sectional profile ($\widehat x-\widehat z$ plane, $y = 20$) 
    of chemical concentration at $t = 500 MCS$ starting from an
    initial set of particles at $x = 75$ with uniform concentration; (A)
    with no diffusion, (B) with diffusion. Particles which move away from
    the plane in the $\widehat y$ direction cannot be seen in the
    figures.}
  \label{fig:chem_profile}
\end{figure}

We check the qualitative agreement of the dispersion coefficient
obtained using our CPM ADE solver for Poiseuille flow along $x$
direction (described in section~\ref{sec:Pois}) in a cylindrical
geometry.  We compare our result for an initial delta distribution of
chemical in the middle of the tube.  After an initial transient, so
that the chemical reaches the boundary in the $y$ and $z$ direction,
we compare the diffusion coefficient of chemical distribution along
the $x$ direction with the analytical result for the effective
diffusion coefficient in Poiseuille flow along the cylinder axis,
\textit{i.e.} $D_{effective} = D_{0} +
\frac{\overline{v}^{2}R^{2}}{48D_{0}} $. Here $D_{0}$ is the diffusion
coefficient in the absence of flow, $\overline{v}$ is the average
velocity and $R$ is the radius of the cylinder.
Figure~\ref{fig:dispersion} shows our results where we have plotted
the variation of effective diffusion coefficient with time (in MCS)
for different mean velocities.  For the analytical results given in
solid lines in fig.~\ref{fig:dispersion}, we use the mean velocity
obtained from the fit of the velocity profile like as shown in
fig.~\ref{profile}.
Figure~\ref{fig:chem_profile} shows
snapshots of chemical profile after $500$ MCS for $D_{0} = 0 \mbox{
  and } D_{0} = 0.12$. In this case, we start with a thin sheet of
chemical in the $x-z$ plane, at $x=75$, \textit{i.e.}, all the particles
at $x=75$ have uniform concentration. We take a cut at $y=20$ to see
the evolved chemical profile. A detailed study of evolution of
chemical profile, effect of embedded objects like sphere under both
stationary and moving condition will be reported in our future
communication.
        
\section{Conclusions}
     
We have implemented fluid flow, advection and diffusion in the
framework of the CPM, avoiding the programming complexity and
computational demands associated with implementing a finite-element or
finite-difference Navier-Stokes simulation and interfacing it with the
CPM lattice.

We have used three biologically relevant test cases to verify our
method.  All our results for diffusion in the presence of boundaries
or moving sources agree very well with corresponding analytical or
finite-element solutions. The errors in our scheme are large if we try
to probe far below the diffusion time scale or the fluid particle length
scale, but the results are qualitatively correct. Thus we must be
cautious when applying this scheme to large $Pe$ number flows.  The
requirement that fluid particles remains connected, limits the method
to low $Re$. However, since most biological mechanisms operate at low
$Re$ our CPM ADE solver is appropriate for many cell-level simulations.


\begin{thebibliography}{99}
\bibitem{physics} C. R. Nugent, W. M. Quarles, and T. H. Solomon, Phys.
  Rev. Lett. \textbf{93}, 218301 (2004); J. D. Seymour, J. P. Gage, S.
  L. Codd, and R. Gerlach, Phys. Rev. Lett.  \textbf{93}, 198103
  (2004); M. Leconte, J. Martin, N. Rakotomalala, and D. Salin, Phys.
  Rev.  Lett. \textbf{90}, 128302 (2003); B. F. Edwards Phys. Rev.
  Lett. \textbf{89}, 104501 (2002);
\bibitem{chem} A. Bancaud, G. Wagner, K. D. Dorfman, J. L. Viovy,
  Anal. Chem.  2004 (submitted);
\bibitem{atmos} T. Yanagita and K. Kaneko, Phys. Rev. Lett.
  \textbf{78}, 4297-4300 (1997).
\bibitem{geo} J. M. Keller, M. L. Brusseau, Environ. Sci. Technol.
  \textbf{37}, 3141 (2003); 
  T. E. McKone and D. H. Bennett, Environ. Sci. Technol. \textbf{37},
  3123 (2003).
\bibitem{disease}G. J. Gibson, C. A. Gilligan, and A. Kleczkowski, Proc.
  R. Soc. Lond. Ser. B. \textbf{266}, 1743-1753 (1999); J. T. Truscott
  and C. A. Gilligan, Porc. Natl. Acad. Sci. U.S.A \textbf{100}, 
9067-9072 (2003).
\bibitem{faber} T. E. Faber, \textit{Fluid Dynamics for Physicists}, 
  (Cambridge University Press, Cambridge 1995).
\bibitem{fem1} P. M. Gresho and R. L. Sani, \textit{Incompressible Flow and the Finite Element Method} , (John Wiley and Sons, 2000); W. Hundsdorfer, J. G. 
  Verwer, \textit{Numerical Solution of
  Time-Dependent Advection-Diffusion-Reaction Equations}, (Springer Series 
in Computational Mathematics, 2003).
\bibitem{lb}Y. H. Qian, D. D'Humieres, and P. Lallemand, Europhys.
  Lett. \textbf{17}, 479 (1992).
\bibitem{flek} E. G. Flekkoy, Phys. Rev. E \textbf{47}, 4247 (1993).
\bibitem{daw} S. P. Dawson, S. Chen, and G. D. Doolean, J. Chem.
  Phys. \textbf{98}, 1514, (1993).
\bibitem{moment}C. P. Lowe and D. Frenkel, Physica A, \textbf{220},
  251, (1995); R. H. Merks, et al, J. of Comp. Phys. \textbf{183},
  563-576 (2002).
\bibitem{ladd} J. C. Anthony, J. Fluid Mech, \textbf{271}, 285-309 1994.
\bibitem{wei} W. Zeng, G. L. Thomas, S. A. Newman, and J.
  A. Glazier, Mathematical Modeling and Computing in Biology and
  Medicine, 5th ESMTB Conference 2002, V.  Capasso editor (Società
  Editrice Esculapio, Bologna, 2003).
\bibitem{psedupod} D. V. Zhelev, A. M. Alteraifi, and D. Chodniewicz,
  Biophys. J. \textbf{87} (2004), 688-695.
\bibitem{adapt} H. C. Berg and P. M. Tedesco, Proc. Natl. Acad. Sci.
  U.  S. A.  \textbf{72}, 3235-3239 (1975).
\bibitem{thrombo} M. H. Kroll, J. D. Hellums, L. V. McIntire, A. L.
  Schafer, and J. L. Moake. Blood.  \textbf{88}, 1525-1541 (1996).
\bibitem{pmn} S. Jadhav and K. Konstantopoulos, Am. J. Physiol.
  Cell Phsiol. \textbf{283}, C1133-C1143 (2002).
\bibitem{genes} M. U. Nollert, N. J. Panaro, and L. V. McIntire,  Ann.
  N. Y. Acad. Sci. \textbf{665}, 94-104 (1992); P. F. Davies, Physiol.
  Rev. \textbf{75}, 519-560 (1995)
\bibitem{roll} E. B. Finger, K. D. Puri, R. Alon, M. B. Lawrence, U.
  H. von Andrian, and T. A. Springer, Nature, \textbf{379}, 266-269
  (1996).
\bibitem{turing} A. M. Turing, Philos.  Trans. R. Soc. London B
  \textbf{237}, 266-269 (1952); A. Taylor, Prog. in Reaction
  Kinetics and Mechanism \textbf{27}, 247-325 (2002).
\bibitem{examples} A D Taylor, S Neelamegham, JD Hellums, CW Smith, and
  SI Simon, Biophys. J. \textbf{71}, 3488-3500 (2004);
\bibitem{fem} M. Pavlin, N. Pavselj, and D. Miklavcic, IEEE Trans.
  Biomed. Engg., \textbf{49}, 605 (2002).
\bibitem{reaction-diffusion} H. G. E. Hentschel, J. A. Glazier and S.A.
  Newman, Proc. Royal Soc. Lond. B. \textbf{271}, 1713 (2004).
\bibitem{ecell} S. Kikichi, K. Fujimoto, N. Kitagawa, \textit{et. al.}, Neural
  Networks, \textbf{16}, 1389 (2003)
\bibitem{alber} M. S. Alber, A. Kiskowski, J. A. Glazier, and Y.
  Jiang, Mathematical Systems Theory in Biology, Communication, and
  Finance \textbf{132}, 1-40 (2003).
\bibitem{first} F. Graner and J. A.  Glazier, Phys. Rev.
  Lett. \textbf{69}, 2013-2016 (1992).
\bibitem{second} J. A. Glazier and F. Graner, Phys. Rev.
  E \textbf{47}, 2128-2154 (1993).
\bibitem{organo} M. S. Alber, M. A. Kiskowski, J. A.
  Glazier, and Y. Jiang, in Mathematical Systems Theory in Biology,
  Communication, and Finance, J. Rosenthal, and D. S. Gilliam, editors
  (IMA 142, Springer-Verlag, New York, 2002), 1-40.
\bibitem{stripe} W. Zeng, G. L. Thomas and J. A. Glazier,
  Physica A \textbf{341}, 482-494 (2004).
\bibitem{morpho} M. Zajac, G. L. Jones, and J. A. Glazier,
  Phys. Rev. Lett. \textbf{85}, 2022 (2000).
\bibitem{stan} S. Maree, \textit{From Pattern Formation to
    Morphogenesis: Multicellular Coordination in Dictyostelium
    discoideum} Ph.D. Thesis (U. Utrecht 2000).
\bibitem{tumor} Emma Stott, N. F. Britton, J. A. Glazier, and M.
  Zajac, Math. and Computer Modelling \textbf{30}, 183-198
  (1999).
\bibitem{roeland} R. M. H. Merks, S. A. Newman and J. A.
  Glazier, Lecture Notes in Computer Science \textbf{3305}, 425-434
  (2004).
\bibitem{limb}  R. Chaturvedi, J. A. Izaguirre, C. Huang, T.
  Cickovski, P. Virtue, G. Thomas, G. Forgacs, M.
  Alber, G. Hentschel, S. A. Newman, and J. A. Glazier, 
  Lecture Notes in Computer Science \textbf{2659}, 39-49 (2003). 
\bibitem{stan} A. F. MarÃee and P. Hogeweg, Proc. Nat. Acad. Sci. USA
  \textbf{98} 3879 (2001).
\bibitem{yi_thesis} Y. Jiang, \textit{Cellular Pattern Formation}, Ph.D.
 dissertation (U. of Notre Dame 1998).
\bibitem{life} E. M. Purcell, Am.  J.  of Phys. \textbf{45}, 3-11
  (1977).
\bibitem{barkema} M. E. J. Newman and G. T. Barkema \textit{Monte Carlo 
Methods in Statistical Physics} (Oxford Univ. Press 1999).    
\bibitem{skai} S. Wong, \textit{A Cursory Study of the Thermodynamic 
 and Mechanical Properties of Monte-Carlo Simulations of the
  Ising Model}, Ph.D. thesis (Notre Dame Univ., 2004). 
\bibitem{murray} J. D. Murray, \textit{Mathematical Biology, Vol 1} (Springer 
Verlag 2002).
\bibitem{landau} E. M. Lifsihtz and L. D. Landau, \textit{Fluid Mechanics}, 
  (Butterworth-Heinemann, 1987).
\bibitem{mimetic}J. Braun and M. Sambridge, Nature \textbf{376}, 655
  (1995).

\end{thebibliography}
\end{document}